\documentclass[preprint, trackchanges]{aastex631} 

\newcommand{\ha}{H$\alpha$}
\newcommand{\sm}{$\sim$}
\newcommand{\kms}{$km~s^{-1}$}
\graphicspath{{./figures/}}

\submitjournal{ApJ}
\shorttitle{Mini-filament by FISS}
\shortauthors{Wang et al.} 
\graphicspath{{figures/}}

\begin{document}
\title{High Resolution Imaging Spectroscopy of a Sigmoidal Mini-filament Eruption}


\author[0000-0001-5099-8209]{Jiasheng Wang}
\affiliation{Institute for Space Weather Sciences, 
            New Jersey Institute of Technology,
            University Heights, Newark, NJ 07102-1982, USA}
\affiliation{Big Bear Solar Observatory,
            New Jersey Institute of Technology, 
            40386 North Shore Lane, Big Bear City, CA 92314-9672, USA}
\affiliation{Center for Solar-Terrestrial Research, 
            New Jersey Institute of Technology, 
            University Heights, Newark, NJ 07102-1982, USA}

\author[0000-0002-5865-7924]{Jeongwoo Lee}
\affiliation{Institute for Space Weather Sciences, 
            New Jersey Institute of Technology,
            University Heights, Newark, NJ 07102-1982, USA}
\affiliation{Big Bear Solar Observatory,
            New Jersey Institute of Technology, 
            40386 North Shore Lane, Big Bear City, CA 92314-9672, USA}
\affiliation{Center for Solar-Terrestrial Research, 
            New Jersey Institute of Technology, 
            University Heights, Newark, NJ 07102-1982, USA}

\author[0000-0002-7073-868X]{Jongchul Chae}
\affiliation{Department of Physics and Astronomy,
            Seoul National University,
            Gwanak-gu, Seoul 08826, Republic of Korea}

\author{Yan Xu}
\affiliation{Institute for Space Weather Sciences, 
            New Jersey Institute of Technology,
            University Heights, Newark, NJ 07102-1982, USA}
\affiliation{Big Bear Solar Observatory,
            New Jersey Institute of Technology, 
            40386 North Shore Lane, Big Bear City, CA 92314-9672, USA}
\affiliation{Center for Solar-Terrestrial Research, 
            New Jersey Institute of Technology, 
            University Heights, Newark, NJ 07102-1982, USA}

\author{Wenda Cao}
\affiliation{Institute for Space Weather Sciences, 
            New Jersey Institute of Technology,
            University Heights, Newark, NJ 07102-1982, USA}
\affiliation{Big Bear Solar Observatory,
            New Jersey Institute of Technology, 
            40386 North Shore Lane, Big Bear City, CA 92314-9672, USA}
\affiliation{Center for Solar-Terrestrial Research, 
            New Jersey Institute of Technology, 
            University Heights, Newark, NJ 07102-1982, USA}

\author[0000-0002-5233-565X]{Haimin Wang}
\affiliation{Institute for Space Weather Sciences, 
            New Jersey Institute of Technology,
            University Heights, Newark, NJ 07102-1982, USA}
\affiliation{Big Bear Solar Observatory,
            New Jersey Institute of Technology, 
            40386 North Shore Lane, Big Bear City, CA 92314-9672, USA}
\affiliation{Center for Solar-Terrestrial Research, 
            New Jersey Institute of Technology, 
            University Heights, Newark, NJ 07102-1982, USA}

\begin{abstract}
Minifilament (MF) eruption producing small jets and micro-flares is regarded as an important source for coronal heating and the solar wind transients through studies mostly based on coronal observations in the extreme ultraviolet (EUV) and X-ray wavelengths. In this study, we focus on the chromospheric plasma diagnostics of a tiny minifilament in quiet Sun located at (71'', 450'') on 2021–08–09 at 19:11 UT observed as part of the ninth encounter of the PSP campaign. Main data obtained are the high cadence, high resolution spectroscopy from the Fast Imaging Solar Spectrograph (FISS) and high-resolution magnetograms from the Near InfraRed Imaging Spectropolarimeter (NIRIS) on the 1.6~m Goode Solar Telescope (GST) at Big Bear Solar Observatory (BBSO). The mini-filament with size \sm1''$\times$5'' and a micro-flare are detected in both the \ha\ line center and SDO/AIA 193, 304~{\AA} images. On the NIRIS magnetogram, we found that the cancellation of a magnetic bipole in the footpoints of the minifilament triggered its eruption in a sigmoidal shape. By inversion of the \ha\ and Ca {\sc ii} spectra under the embedded cloud model, we found a temperature increase of 3,800 K in the brightening  region, associated with rising speed average of MF increased by 18~\kms. This cool plasma is also found in the EUV images. We estimate  the kinetic energy change of the rising filament as 1.50$\times$$10^{25}$~ergs, and thermal energy accumulation in the MF, 1.44$\times$$10^{25}$~ergs. From the photospheric magnetograms, we find the magnetic energy change is 1.57$\times$$10^{26}$~ergs across the PIL of converging opposite magnetic elements, which amounts to the energy release in the chromosphere in this smallest two-ribbon flare ever observed.

\end{abstract}


\section{Introduction}\label{sec:intro}


Solar filaments are typically dense, cool plasma formulated in chromosphere along the polarity inversion line (PIL) during flux emergence \citep[e.g.,][]{1967SoPh....2..451B, 1971SoPh...19...59F, 1985SoPh..100..397Z, 1989ApJ...345..584S, 2017ApJ...836...63T}. The arch filament enveloped in the emerging magnetic bipole expands as it rises from the solar surface. 
The filament erupts as its configuration with surrounding fields is destabilized by reconnection in different scenarios, such as breakout, kink instability, and tether-cutting \citep{2015ApJ...805...48B, 2019ApJ...885L..15K, 2022ApJ...932L..18K}. 
Solar filament eruptions are often associated with a wide range of dynamic solar activities from as large as coronal mass ejections (CMEs) to small-scale spicule activities by which free energy is transported to the corona from the chromosphere. 
In active regions, the sigmoid configurations, which are defined as S-shaped loop arcades of non-potential magnetic field, have typical length of 100$\arcsec$. Non-potentiality of the kinked magnetic field flux ropes tend to drive eruption in the PIL (polarity inversion lines) of the sigmoidal regions by tether-cutting reconnection and form arcade loops \citep{2010ApJ...725L..84L, 2012ApJ...745L...4L}. 
Strong non-potential bipolar magnetic configuration is also identified in X-ray and EUV micro-sigmoids \citep{2005A&A...434..725M, 2010ApJ...718..981R, 2013MNRAS.431.1359Z}. 

Mini-filaments are scale-down version of solar filaments, which extend less than 25 Mm and last for about 50 min in \ha\ images \citep{Wang_2000}. Similar to large-scale filaments, loop-like topology of mini-filaments usually lie above the photospheric polarity inversion line (PIL) of small-scale adjacent opposite-polarity magnetic fields, which are distinguished from ``surge-like macrospicules'' that emanating from network field concentrations on the same size scale. 
The earliest systematic studies of small-scale filaments utilizing \ha\ observations from Big Bear Solar Observatory (BBSO) by \cite{Hermans_1986} discovered elongated dark features and characterized that: 1) With occurrence rate of 600 per 24 hour day, the average duration of eruptive phase was 26 minutes in total lifetime of 70 minutes; 2) Majority of events are spatially related to cancelling magnetic features; 3) Small-scale filament share similarities with large scale filaments that often has one or both ends terminate at cancellation site. 
The total mass and kinetic energy are estimated to be 10$^{13}$~g and 10$^{25}$~ergs, respectively.  

The eruptions of mini-filaments are also found to often produce small scale coronal mass ejections, coronal jets, and microflares \citep{Innes_2009,Raouafi2016}, which is thus of interest to the recent topic a small solar eruption that could contribute correspondingly small transients to the coronal heating and solar wind acceleration.
Observations of the roots of coronal heating in chromosphere proved that it cannot be driven solely by energy deposition in the coronal heights; instead, mass and energy supply from lower atmosphere are required through ubiquitous upflows across the magnetic configuration in the heating process of corona \citep{DePontieu2009,Tian2017}. 
The ejected cold material from filament eruptions in chromosphere called surges have been observed in \ha\ filtergrams \citep{Harrison1988, Schmieder1995, Chen2022, Schmieder2022} with bright/flaring points separating from footpoints of the surge during onset of eruptions. 
The brightening primarily observed in soft X-ray (SXR) wavelengths is located where the \ha\ surges originate, in which magnetic reconnection between twisted cool loops and open field lines leads to the formation of surges that coincide with X-ray subflares \citep{Shibata1992,Alexander1999}. Coronal counterparts of the erupted filament, jets, are observed in extreme ultraviolet filtergrams \citep[EUV,][]{Wang1998,Li2023}, as well as at SXR \citep{Innes2016, Sterling2022}. Observational facts of the scenario that magnetic reconnection bewteen confined flux loop and overlying field in different wavelengths suggest the formation of cool and hot ejecting components from minifilament eruptions \citep{Shen2017}. Coronal bright points (CBPs) of network field configuration of opposite polarities are found to be heated to their brightness temperature maxima with occurrence of minifilament eruptions \citep{Hong2014}.

A schematic model of MFs eruptions that has characteristics different from standard jet model \citep{Shibata1992} was proposed by \cite{2010ApJ...720..757M}, in which the sheared core field carrying cool plasma to begin erupting when X-ray brightening in the base arch is observed. Such dynamic small-scale filament eruptions are often observed to initiate blowout jet scenario through interchange reconnection at the boundary of open fields and closed flux loops \citep{Shen2012,Adams2014,Li2015}, and show obvious asymmetric brightening in the loop called jet bright points (JBPs) \citep{Sterling2015}. The external reconnection is driven by flux cancellation \citep{Panesar2017,McGlasson2019,Sterling2020} at the polarity inversion line and opens filament field to allow plasma ejection through the undergoing internal reconnection \citep{Shen2012, Adams2014, Panesar2016,Panesar2018}. Recently, by reviewing jet producing minifilament events and flux cancellation model, \cite{Sterling2020} proposed a possible mechanism for magnetic switchbacks in the solar wind detected by Parker Solar Probe (PSP), in which the jetlets carry twist from flux cancellation in the interplanetary propagation. In-situ observations of solar wind show expansion of scale and duration small-scale magnetic flux ropes over propagation \citep{Chen2020}, which inversely, are used to investigate their possible solar sources \citep{Bale2021,Lee2022}. The occurrence rate of small-scale solar eruptions 0.02 per unit area per hour is arguably sufficient to heat corona and generate Alv{\'e}nic pulses for switchbacks \citep{Wang2022,Raouafi2023}.



The questions raised to be addressed are: How do small-scale filaments evolve in the quiet Sun? What is the energy distribution in the chromosphere and corona in the event of MF eruption? How much free energy from underlying photospheric magnetic fields is released by the coronal heating and solar wind acceleration? With high-resolution \ha\ and Ca II spectra, this study focus on the above questions by investigating a newly observed two-ribbon microflare associated with partial minifilament eruption.


In this paper, we study spectroscopic and dynamic properties of a mini-filament observed with the \ha\ and Ca {\sc ii} lines offered by FISS together with EUV wavelengths (AIA) images. 
Although significant progress has been made in erupting minifilaments in X-ray or EUV wavelengths, earlier identification of small scale eruption in \ha\ wavelength minifilaments \citep{Wang2000} suggests that fact that jets may have both hot and cool components. In this regard, spectral diagnostics available from the chromospheric lines will complement the existing results obtained from X-ray/EUV wavelengths. We focus on the spectral diagnostics of the the chromospheric temperature and density variation around the time of the minifilament eruption along with underlying magnetic field detected by the GST/NIRIS to provide insight into small-scale solar eruption along with jet-formation.

\begin{figure}[ht!]
    \centering
    \includegraphics[width=\linewidth]{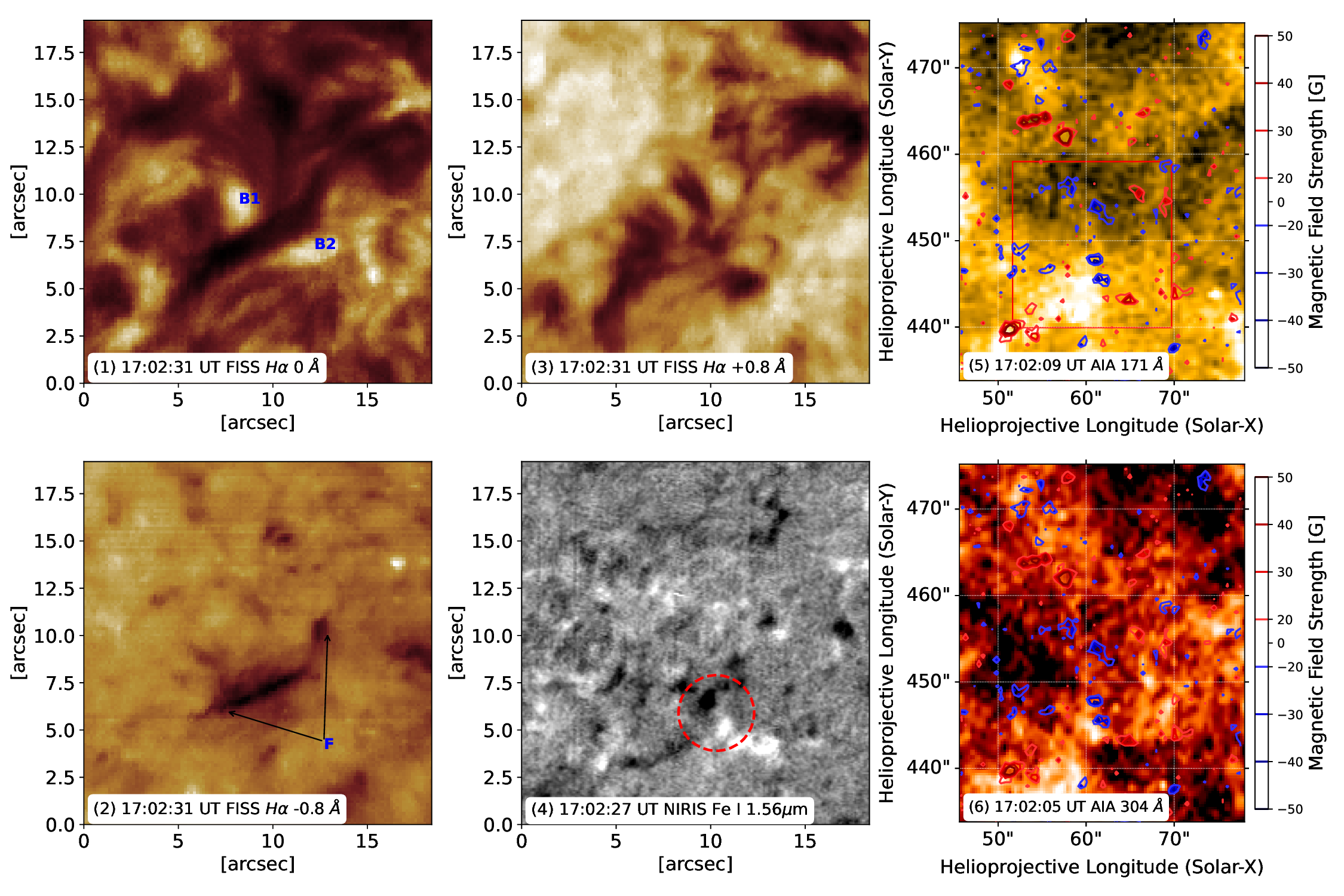}
    \caption{
    Minifilament eruption and associated brightenings. \ha\ images in the line center (1) and wings at $\pm$0.8~\AA\ (2, 3) show the rising MF (labled F) and associated with brightening (B1 and B2) on the sides of the filament channel. The high-resolution magnetogram (4) shows the magnetic bipole associated with the brightening (a dashed circle). SDO/AIA images in the 171~\AA\ and 304~\AA\ channels (5, 6) show the brightening associated with the erupting filament. The superimposed contours are HMI magnetogram at $\pm$20, $\pm$30, $\pm$40, and $\pm$50~G. The red box in (5) denotes the FoV of (1–4).}
\label{fig:f1}
\end{figure}

\section{Observations and Data Analysis} \label{sec:data}
\subsection{Quiet Sun data}
On 2021 August 7, BBSO/GST provided ground-based support to Parker Solar Probe (PSP) observations with multi-wavelength instruments. GST observations of quiet Sun filament evolution were conducted on the target that is located at (71$\arcsec$, 450$\arcsec$) before the ninth encounter of the PSP campaign passing through perihelion (2021--08--09 19:11~UT) on the far side. We utilized 304~\AA\ and 1600~\AA\ images taken by the Atmospheric Imaging Assembly \citep[AIA;][]{2012SoPh..275...17L} on board Solar Dynamic Observatory (SDO) to observe the brightening associated with the MF eruprion in UV/EUV channels.

High-resolution observations were carried out by the 1.6~m Goode Solar Telescope \citep[GST;][]{2010AN....331..636C} in Big Bear Solar Observatory (BBSO), with the advantage of a high-order adaptive optics system combining 308 sub-apertures and reconstruction technique for solar speckle
interferometric data \citep{2008A&A...488..375W}. Photospheric optical continuum images are obtained by Broadband Filter Imager (BFI) with a TiO ﬁlter (705.7~nm; 10~\AA\ band-pass). Spectroscopic polarization measurement of the Fe I 1565~nm line (0.25~\AA\ band-pass) is taken by NIRIS with a round FOV of 80$\arcsec$ at 0$\farcs$24 resolution and 42~s cadence.

\ha\ observation is achieved with Fast Imaging Solar Spectrograph (FISS), which is capable of recording dual-band Echelle spectrograph at \ha\ line and Ca {\sc ii} 854.2~nm simultaneously. 
Only the \ha\ line data are used in the present work. Each \ha\ spectrogram (slit image) consists of 256 spectral profiles that represent the spatial variation along the slit with a sampling of 0$\farcs$16, and each spectral profile, of 512 data points that cover 0.97~nm (9.7~\AA) with a sampling of 1.9~pm (19~m\AA). The position of the slit on the image plane was successively shifted with a step size of 0$\farcs$16, and one raster scan comprising 100 steps covered a 32$\arcsec$~$\times$~41$\arcsec$ area inside the field of view. The raster scan was repeated every 22~s for half an hour.

\subsection{Reduction and model fitting} \label{sec:observation}

To process the FISS spectrographs, an average of the observed background solar region is taken as a reference profile. All the profiles are normalized by the maximum intensity (a proxy of the continuum intensity) of the reference profile. The data profiles are then corrected to the reference profile by a multiplication factor that is close to unity so that contrast in far wings vanishes while maintaining contrast accuracy in the \ha\ line center. The stray light effect  in the observation is removed by following the calibration process of \cite{2013SoPh..288....1C}. Thermodynamic properties (i.e., temperature and line-of-sight velocity) of the filaments are obtained from \ha\ spectrum inversion. In this study, we applied the embedded cloud model proposed by \cite{2014ApJ...780..109C} to the calibrated data, which assumes low-lying cloud features (e.g., minifilaments, spicules) are partially embedded in the chromosphere that displays both emission and absorption, then use regression method to the free parameters to fit model with observation. According to Beckers' cloud model, the absorption profile is uniform inside the feature and subject to Doppler broadening only, giving the optical depth 
\[ \tau_{\lambda} = \tau_{0} \exp[-(\frac{\lambda-\lambda_{1}}{W})^{2}] , \] 
and radiative transfer function
\[ I_{\lambda,obs} = I_{\lambda,in} \exp[-\tau_{\lambda}] + S[1 - \exp(-\tau_{\lambda})] , \]
where $I_{\lambda,obs}$ and $I_{\lambda,in}$ represents observed and incident intensity of light; parameters $\tau_{0}$, $S$, $W$, and $\lambda_{1}$ are optical depth at peak absorption, source function, Doppler width and peak absorption wavelength at line-of sight (LOS) velocity $v$.  
By assuming incident intensity (I$_{\lambda,in}$) to equals intensity of light (R$_{\lambda,in}$) at the same height in chromosphere, basic radiative transfer function is reduced to 
\[ R_{\lambda,obs} = R_{\lambda,in} \exp(-t_{\lambda}) + s[1 - \exp(-t_{\lambda})] ,\]
in which the $R_{\lambda,obs}$, $R_{\lambda,in}$, and $s$ are ensemble-average quantities of observed reference profile, intensity of light, and source function, respectively. $t_{\lambda}$ is defined as 
\[ t_{\lambda} = t_{0} \exp[-(\frac{\lambda-\lambda_{2}}{\omega})^{2}], \]
with three introduced parameters $t_{0}$, $\lambda_{2}$, and $\omega$. 
By resolving the previously defined parameters\citep[also see,][]{1968SoPh....3..367B,2007ASPC..368..217T,2010MmSAI..81..769B}, we obtain temperature and LOS velocity from the equations
\[ v = \frac{\lambda_{1} - \lambda_{0}}{\lambda_{0}}c , \]
and 
\[ W = \frac{\lambda_{0}}{c} \sqrt{\frac{2KT}{M} + \xi^{2}} \]
Combining spatial and wavelength information of \ha\ and Ca II spectra taken by FISS, we are able to derive thermal and dynamic properties of the chromospheric structures.
Further, with multilayer spectral inverison (MLSI) method developed by \cite{Chen2020}, which divides the relatively wide range of atmosphere into finite number of layers consisting of photosphere, lower and upper chromosphere, we estimate and compare kinetic properties in different height during the eruption of MF.   
\section{Results} \label{sec:results}
An event of minifilament eruption in multiple channels (FISS \ha\ lines, AIA 171\AA, 304\AA, and NIRIS magnetograms) is presented in Figure \ref{fig:f1}. The x-axis is along East-West direction in all panels. Figure \ref{fig:f1}(1)--(4) display the GST observations of the MF evolution from spicular bundles between opposite polarities of the same FOV, which is indicated by the red box in panel 5. The MF in sigmoid configuration is displayed in \ha\ filtergrams at central line and $\pm$0.8~\AA\ off at blue wing (see panel 1--3). Two parts of the arcade loops coincides at [10$\arcsec$, 7.5$\arcsec$], with brightening structures B1 and B2 appearing across two sides of the central region of the sigmoid, and fan-spined dark features observed with \ha\ at $-0.8$~\AA\ wavelength. The magnetic bipole at the MF footpoints are circled in panel 4. Upper chromospheric and Coronal counterparts of the MF at higher temperature is observed in AIA 304~\AA (50000~K) and 193~\AA (1.25~MK) (see Figure \ref{fig:f1}(5) and (6)), respectively. Manifestation of a miniature version of two-ribbon flare is observed in a supplementary \ha\ and AIA (193~\AA\ and 304~\AA)movie. Contours of longitudinal magnetic field at $\pm$50~G are superimposed on AIA 304~\AA\ and 193~\AA, with positive (negative) field represented in red (blue). The polarity inversion line lies along the direction of MF, which is between opposite polarities of strong magnetic elements.

\begin{figure}[ht!]
    \centering
    \includegraphics[width=\linewidth]{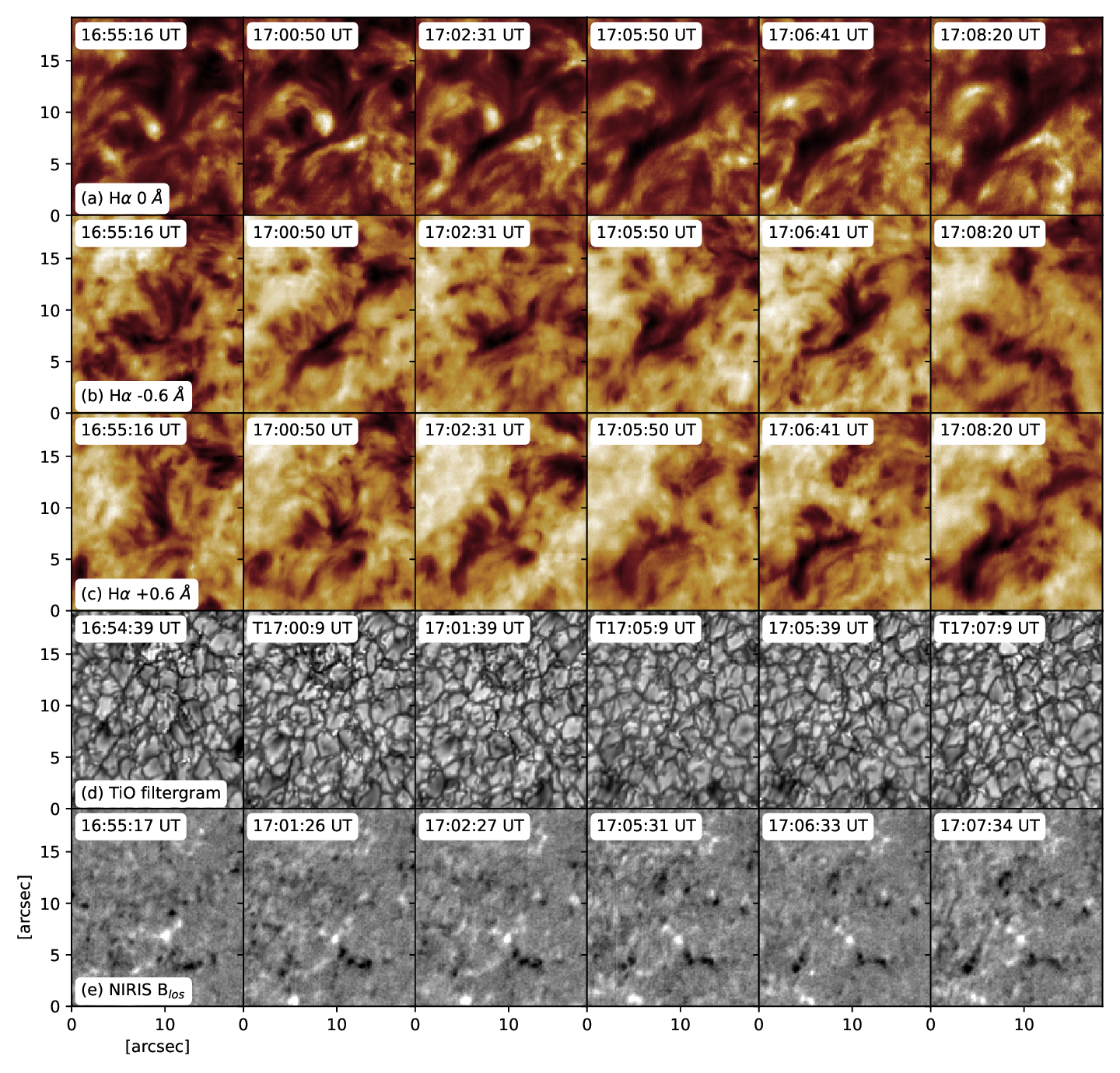}
    \caption{Evolution of minifilament in \ha\ spectrographs, Broadband TiO filtergrams, and magnetograms (top to bottom) from 16:55~UT to 17:08~UT. (a), (b) and (c) show snapshots of the MF eruption in \ha\ images at line center and $\pm$0.6 off center wavelengths, respectively. (d) show granulation underneath the MF in photosphere in broadband TiO images. (e) show LOS magnetograms of the MF event. }
    \label{fig:f3}
\end{figure}

Detailed eruption process of the MF is shown in Figure \ref{fig:f3}. Footpoints of the emerging MF was preoccupied by spicule bundles surrounding the bright point as seen in \ha\ line-center wavelength at 16:55:16~UT. The MF emerged from the bright points with one footpoint anchored in place, of which the rising flux formed two conjugated arcade loop that extend in opposite directions at 17:00:50~UT. The MF upflow is shown in the \ha$-$0.8~\AA\ at the same time as appearance of MF sigmoid structure. The eruptive activity of the rising sigmoid is observed at 17:05:50~UT with bright points seperating from the footpoints, followed by prominent downflow three minutes later as seen in \ha$+$0.8~\AA, as the rising MF vanishes.

\begin{figure}[ht!]
    \centering
    \includegraphics{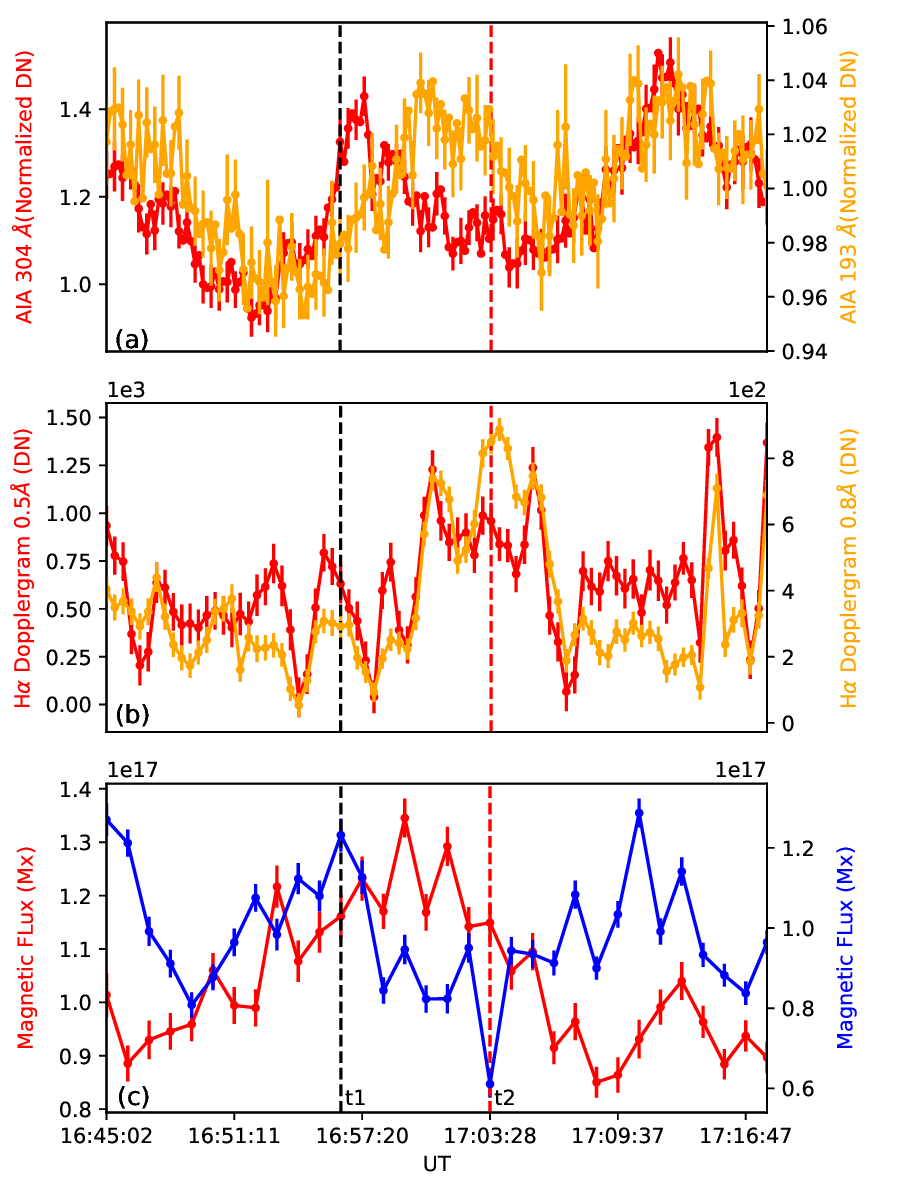}
    \caption{Time profile of the MF event in EUV, \ha\ wavelengths, and magnetiograms. Red and yellow curves in (a) show EUV intensity of the minifilament core region in 304~\AA\ and 193~\AA\ channels. (b) shows changes of normalized \ha\ intensity at $-$0.5~\AA\ and $-$0.8~\AA\ from absorption line center. (c) shows changes of positive (red) and negative (blue) magnetic flux of the magnetic bipole corresponding the brightening in B1 and B2 in Fig.\ref{fig:f1}. Black and red dashed lines mark the time of minifilament activation ($t_1$) in \ha\ center line and the start time of minifilament eruption ($t_2$). The errorbars in each profile indicate estimation of errors at 3-$\sigma$ deviation.}
    \label{fig:ov_profile}
\end{figure}

Figure \ref{fig:ov_profile} show time variation of the MF intensity in EUV wavelengths by AIA 193 and 304~\AA\ channels(Fig.\ref{fig:ov_profile}(a)). After subtracting background oscillation, we found the activation and rising of MF ($t_1$) is followed by EUV dimming in 304~\AA, and intensity starts to increase when start of the MF eruption ($t_2$) is observed in \ha\ images (Fig.\ref{fig:f3}(a-c)). And AIA 193~\AA\ profile shows similar trend, while the intensity reaches maximum \sm2~minutes later than in 304~\AA\ before dimming between $t_1$ and $t_2$. The \ha\ profiles (Fig.\ref{fig:f3}(b) in both $-$0.5 and $-$0.8~\AA\ show increase of Doppler blue-shift indicated by absorption contrast starting from $t_1$ and decrease to background level \sm3~minutes after $t_2$. Fig.\ref{fig:f3}(c) show that positive magnetic flux decreased by 6.2$\times$$10^{16}$~Mx through MF eruption while the negative magnetic flux decreased by 3.0$\times$$10^{16}$~Mx from $t_1$ to 3~minutes after $t_2$. 

\begin{figure}
    \centering
    \includegraphics{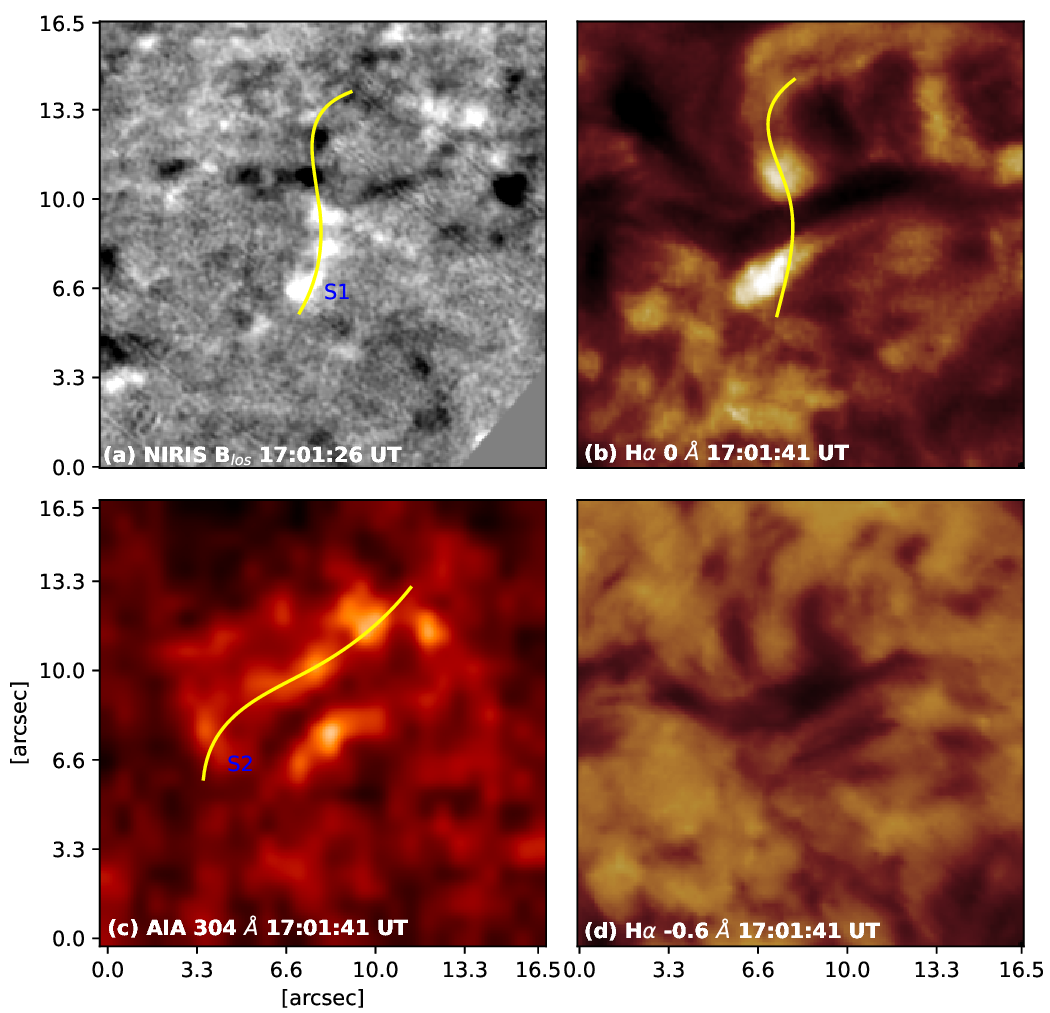}
    \caption{Images of MF eruptions at different wavelengths. (a) shows magnetic bipole at footpoint of the MF sigmoid at (positive) flux peak time. (b) and (d) show the brightenings at footpoint marked (as B1 and B2) in Fig.\ref{fig:f1}(a) and the MF in \ha\ at line center and $-$0.6~\AA. (c) shows the flaring ribbon along the the MF in AIA 304~\AA\. The yellow curves in (a) and (b) denote slit \textbf{S1} across the MF, and the yellow curve in (c) denotes the slit along the MF sigmoid.}
    \label{fig:map_s1s2}
\end{figure}

\begin{figure}
    \centering
    \includegraphics{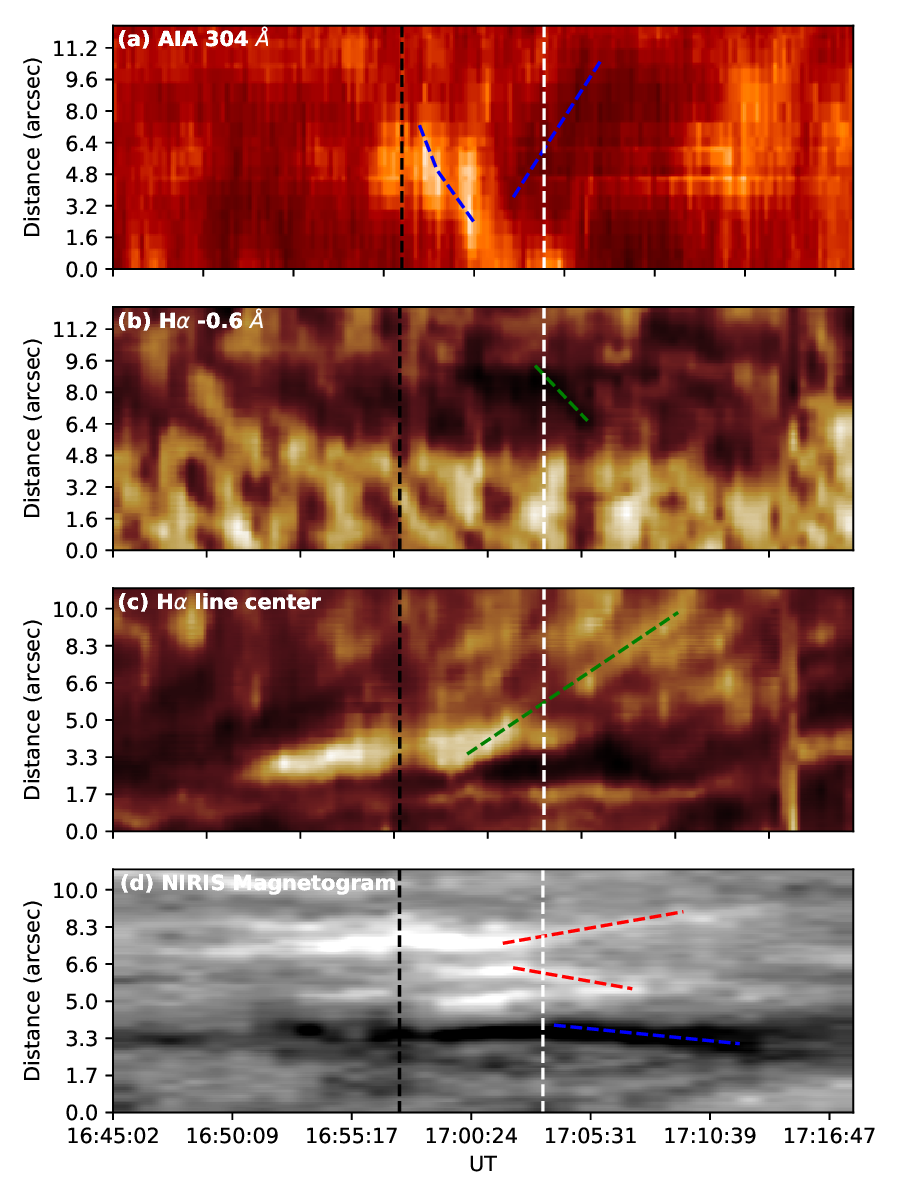}
    \caption{Time-distance diagrams of the minifilament and associated bright ribbons. (a) and (b) show intensity variation of EUV wavelength at 304~\AA\ from AIA and \ha\ $-$0.5~\AA\ off line center, along the slit \textbf{S2}, respectively. Blue and green dashed lines during 16:57~UT--17:00~UT and 17:03~UT--17:05~UT indicate ribbon and MF motion, respectively.(c) and (d) show flaring ribbon motion diverging from the MF and divergence of magnetic bipoles at the MF footpoint along the slit \textbf{S1} across the MF. Green dashed line indicates the motion of bright ribbon front. Red (blue) dashed lines indicate positive (negative) magnetic flux motions. Black and white dashed lines mark the time of minifilament activation ($t_1$ in fig.\ref{fig:ov_profile}(c)) in \ha\ center line and the start time of minifilament eruption ($t_2$ in fig.\ref{fig:ov_profile}(c)) .}
    \label{fig:slitmaps}
\end{figure}

Two slits \textbf{S1} and \textbf{S2} are created along the magnetic bipole direction and along the MF sigmoid trace, respectively (see Figure \ref{fig:map_s1s2}(a) and (c)). Figure \ref{fig:slitmaps} show time-distance diagrams of the MF along \textbf{S1} (Fig.\ref{fig:slitmaps} (a) and (b)) and \textbf{S2}(Fig.\ref{fig:slitmaps} (c) and (d)). In Fig.\ref{fig:slitmaps} (a), we found the motion of the bright point and the following dark feature along the MF sigmoid. The apparent speed of bright point is 38--54~\kms, and the speed of dark feature seen in Fig.\ref{fig:slitmaps} (a) and (b) is 21.5$\pm$1.0~\kms. Fig.\ref{fig:slitmaps}(c) shows the ribbon propagation from footpoint of the MF sigmoid at a speed of 11$\pm$0.8~\kms. Transverse motions of the magnetic bipole footpoints in Fig.\ref{fig:slitmaps}(d) show speed of 1.9$\pm$0.5~\kms. 

In the magnetic field evolution the negative polarity is stationary in position while the positive polarities emerge and migrate towards negative polarity in the rising phase of MF. Field strengths in both positive and negative polarities decrease by \sm30$\%$ (24$\%$) in MF eruption (see Figure \ref{fig:ov_profile}(c)).
With accurate co-alignment in \ha\ images and AIA images (193~\AA, 304~\AA), we distinguished this MF eruption with FISS spectra and the exact location of its coronal counterparts in AIA EUV images. This enables us to study thermodynamic properties of the MF eruption from characteristics of the spectra and their connection with coronal dynamics. We focus our spectral analysis of \ha in the following subsection, on the transient part of the MF which survived \sm10~minutes in \ha\ images.


\subsection{Spectrosopic Analysis of the Minifilament}\label{sub1}
\begin{figure}
    \centering
    \includegraphics[width=\linewidth]{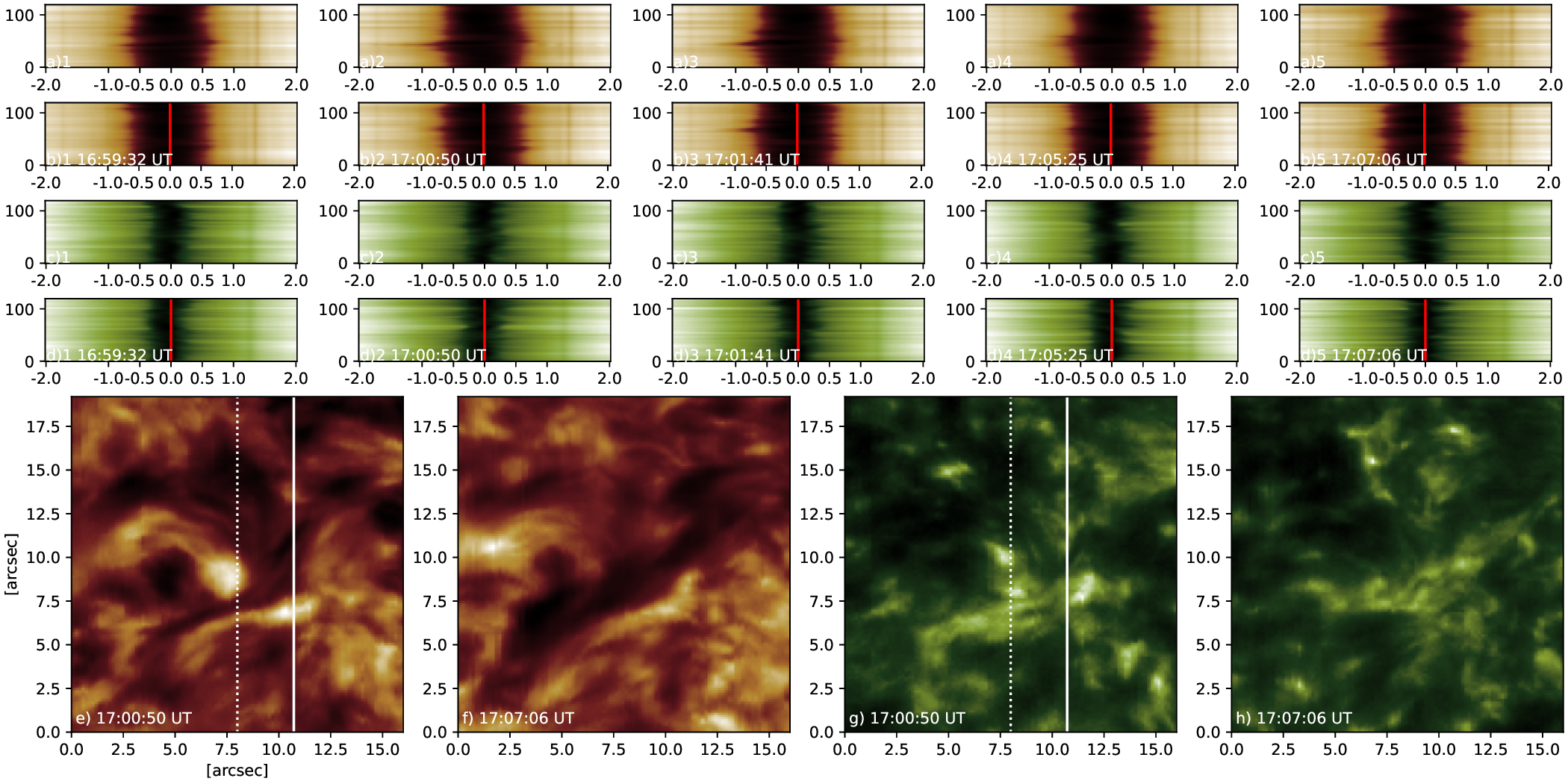}
    \caption{Panels (a)-(d): FISS spectra of \ha ((a) and (b)) and Ca II 8542\AA\ ((c) and (d)) lines in time sequence from rise of the MF to eruption. Spectra ((a) and (c)) of the dashed slit position pass through the ML loop and a bright kernel; Spectra ((b) and (d) of the solid slit position pass through an extending bright ribbon. Panel (e)-(h): Raster images of \ha ((e) and (f)) and Ca II 8542~\AA\ ((g) and (h)) at 17:00:50~UT and 17:07:06~UT.}
    \label{fig:spectra}
\end{figure}
We show the spectroscopic data obtained with 22~s cadence between 16:59:33~UT and 17:07:06~UT in Figure \ref{fig:spectra}. The raster images of \ha\ and Ca II 8542~\AA\ are taken at line center(see Figure \ref{fig:spectra}(e-h)), respectively. The \ha\ images display an $\epsilon$-shape MF structure near the center of the FOV, with two J-shaped arms reaching \sm5\arcsec\ from the junction (\sm[8\farcs0, 7\farcs5]). We select a slit position across the prominent rising arm of MF structure as well as bright point in vicinity. In the \ha\ spectral profiles, strong blue-shifted absorption is detected from 17:00:50--17:02:30~UT. The Ca II 8542~\AA\ spectral profiles show narrowed absorption in the line center, with corresponding brightening fibrils (\sm[8\farcs0, 7\farcs0]) in raster images (Figure \ref{fig:spectra}(g)). Slit 2 is selected across the bright ribbon along the rising segment of the MF and cut through the J-shaped arm at \sm[10\farcs5, 10\farcs0]. At 17:07:06~UT, The MF feature in low opacity broaden in northwest direction, with Ca II 8542~\AA\ bright fibrils ceding underneath the MF loop. 

\begin{figure}[ht!]
  \centering
  \includegraphics[width=\textwidth]{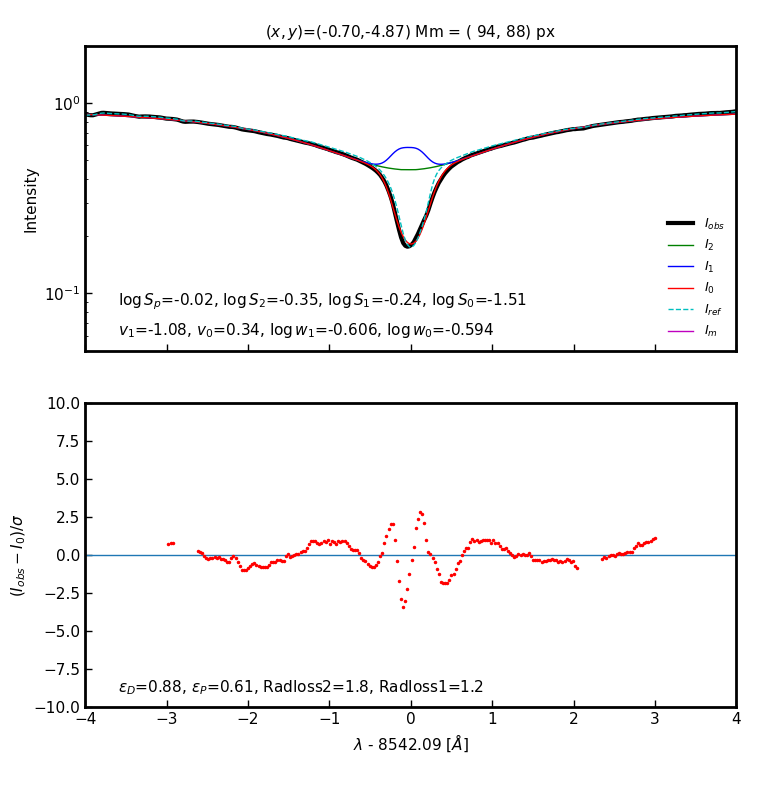}  
  \caption{}
  \label{fig:f9_1}
  \end{figure}
  
\begin{figure}
  \centering
  \includegraphics[width=\textwidth]{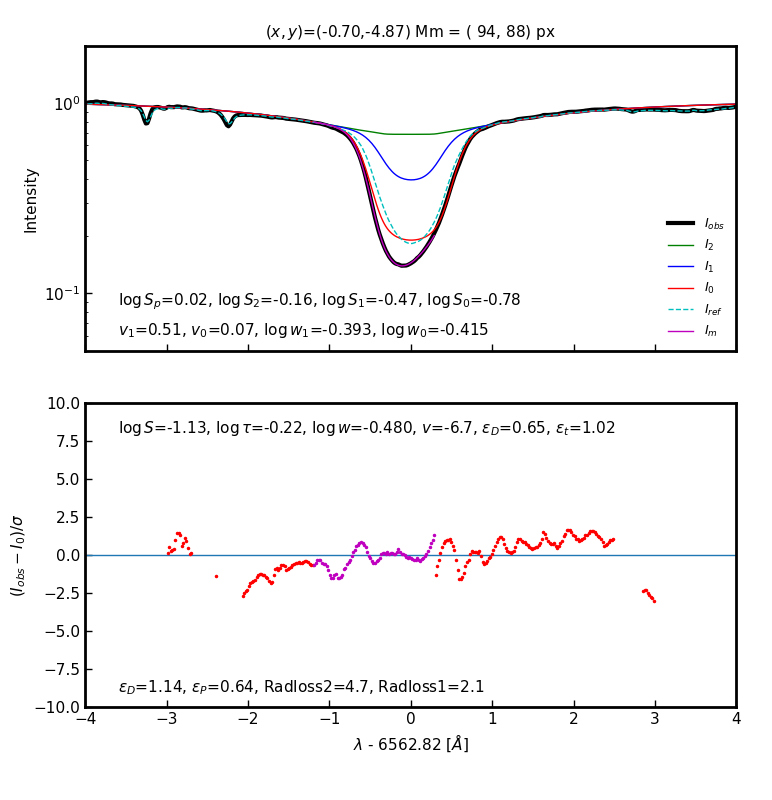}  
  \caption{MLSI results of (a) Ca II and (b) \ha\ line spectra in the mini-filament. The layers of the spectral fitting are denoted as 0 (top layer), 1 (middle layer), and 2 (bottom layer) in the chromospheric heights. Black curves indicate observed spectra of selected feature point F. Reference profile is represent in dotted cyan curve, and fitting results are represented in red, blue and green curves from top to bottom layers, respectively.}
  \label{fig:f9_2}
\end{figure}


We applied the most update MLSI code to \ha\ and Ca II 8542~\AA\ spectra at different points (B,F, and R) in the slit postions displayed in Figure \ref{fig:spectra}. Given the source function and opacity are variable along chromosheric height, MLSI generate a composite of spectral parameters to represent different layers. In Figure \ref{fig:f9_1} and figure \ref{fig:f9_2},  for the spectra inside the MF, the \ha\ profile fitting is based on the 4-layer model fit of the profile. Spectral wavelengths in the cloud feature range [-1.2~\AA\, 0.30~\AA] has been fitted by the 4-layer model, while is fitted by 3-layer model outside the wavelengths of the cloud regime, in which, the top layer is like a cloud. The MF is characterized by source function of \sm0.08,  optical depth of 0.6,  Doppler width of 0.33~\AA, and Doppler velocity of $-$6.7~\kms. The temperature of the MF is around 1$\times$$10^{4}$~K. The Doppler width of the \ha\ in combination with the Ca II 8542~\AA\ Doppler width of 0.25~\AA\ yields the temperature estimate of 9,400~K at the beginning of the MF rising phase. The above values may vary depending on the choice of the cloud wavelength range, but not much under the requirement of good fitting. 
With above selection of fitting parameters, we have verified that at onset of the filament rising phase, the MF can be treated as non-emitting cloud. Therefore, in the following results of spectral inversion, we focus on the central regime of the spectral wavelengths using parameters of the cloud features. 

\begin{figure}[ht!]
    \centering
    \includegraphics[width=\textwidth]{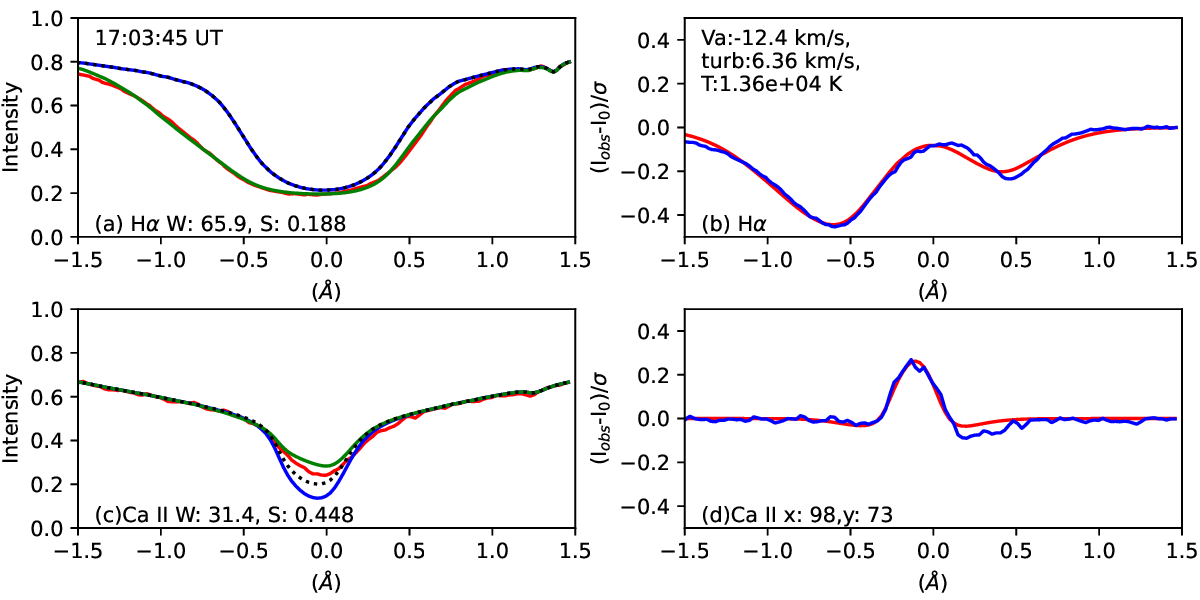}
    \caption{Absorption profiles of the cloud feature at start of the MF rising phase (17:00:50~UT). The blue curves and red curves represent background profiles and observation data in the absorption line profiles, respectively. The green curves represent fitting of cloud features of the MF in range $-$1.5 -- 1.5~\AA. In the contrast profiles, blue and red curves represent data reduction and fitting of cloud features, respectively.}
    \label{fig:f10}
\end{figure}

Figure \ref{fig:f10} show inversion of \ha\ and Ca II spectral profiles using embedded cloud model (ECM) at start time (17:00:50~UT) of the MF eruption. The strongest upflow motion indicated by the blue shift of \ha\ is observed at this moment. The Doppler velocity is estimated reaching \sm18~\kms\ by using lambdameter method \citep{2013SoPh..288...89C} on \ha\ profile. In Ca II 8542~\AA\ profile, we find the enhanced intensity with redshift, indicating bright fibrils associated with the MF rising. We interpret such relation as photospheric magnetic field cancellation driving embedded filament upward motion.  

   \begin{figure}
    \centering
        \includegraphics[width=1\linewidth]{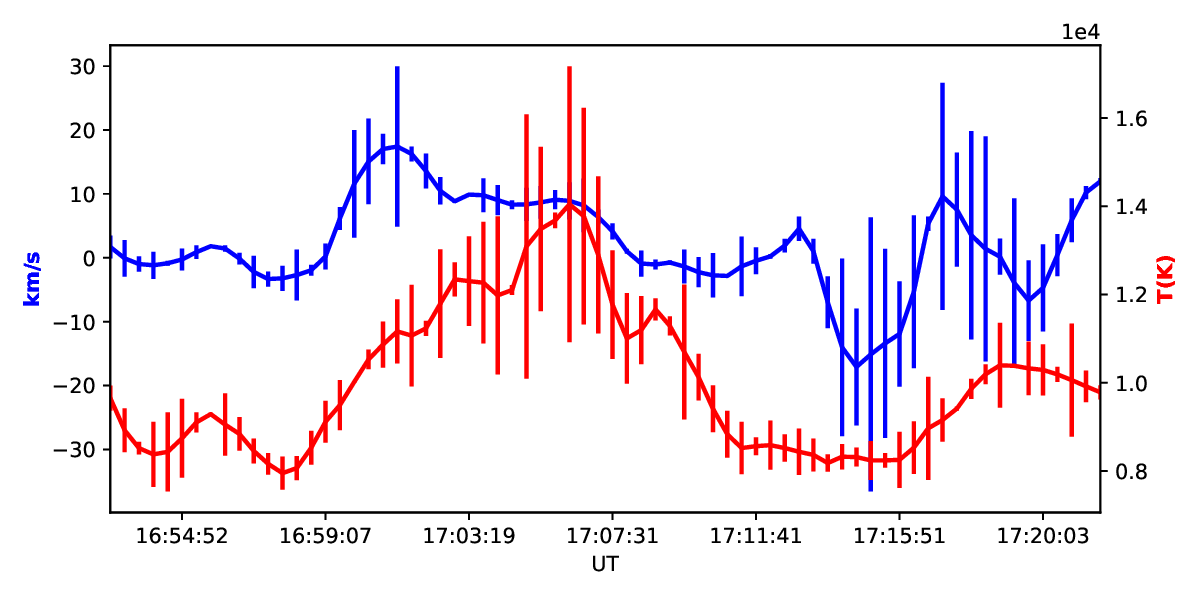}
        \caption{}
        \label{fig:Ng1} 
	\end{figure}
	
	\begin{figure}
        \includegraphics[width=1\linewidth]{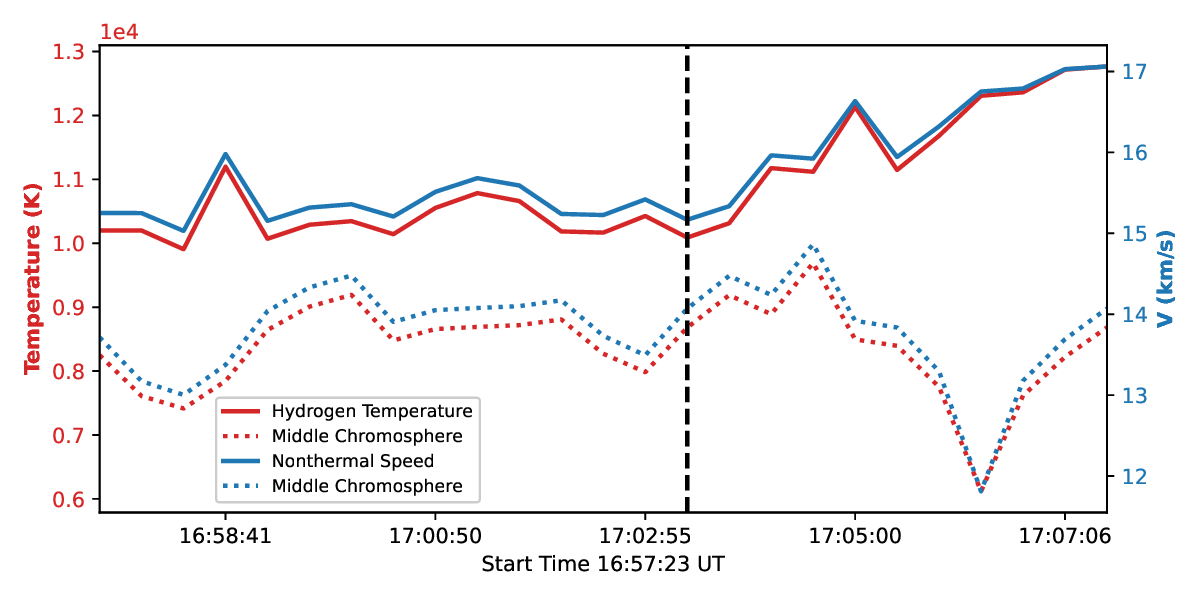}
        \caption{Temperature and Doppler velocity of the MF. In the middle chromosphere temperature decreased by 2500~K in 4~minutes after eruption while in the higher chromosphere temperature continuously rise until reaching 12,600~K in 7~minutes. The solid (dashed) red and blue curves represent temperature and nonthermal speed in the high (middle) chromosphere, respectively.}
        \label{fig:Ng2}
 	\end{figure}

Temperature variation and Doppler velocity change of the MF before and after eruption is shown in figure \ref{fig:Ng1} and figure \ref{fig:Ng2} from 16:56:58~UT to 17:12:56~UT. As shown in the velocity profile, the MF has prominent upflow from the rising phase (17:00:50~UT) to eruption (17:07:06~UT).  Peak value 28~\kms\ of Doppler velocity is reached when the dark feature become visible in \ha\ blue-wing images (see Figure \ref{fig:f3} (b2)). Temperature of cloud feature show continuous increase in 403~s from MF rise to eruption, followed by an abrupt decrease in the next 160~s. Temperature reaches maximum at 17:06:15~UT when the eruption occurs and filament structure eject laterally (see Figure \ref{fig:f3} (b4) and (b5)). The acceleration in upflow velocity and the photospheric flux cancellation show close association, during which the magnetic flux reaches minimum at the same as the upflow speed reaches maximum. However, time discrepancy is found between temperature rise of the MF feature and flux cancellation but is co-temporal with the following magnetic flux increase. We interpret this relation as continuous emergence of flux bring the MF to upper chromospheric heights through reconnection driven acceleration. 

\begin{figure}
    \centering
    \includegraphics{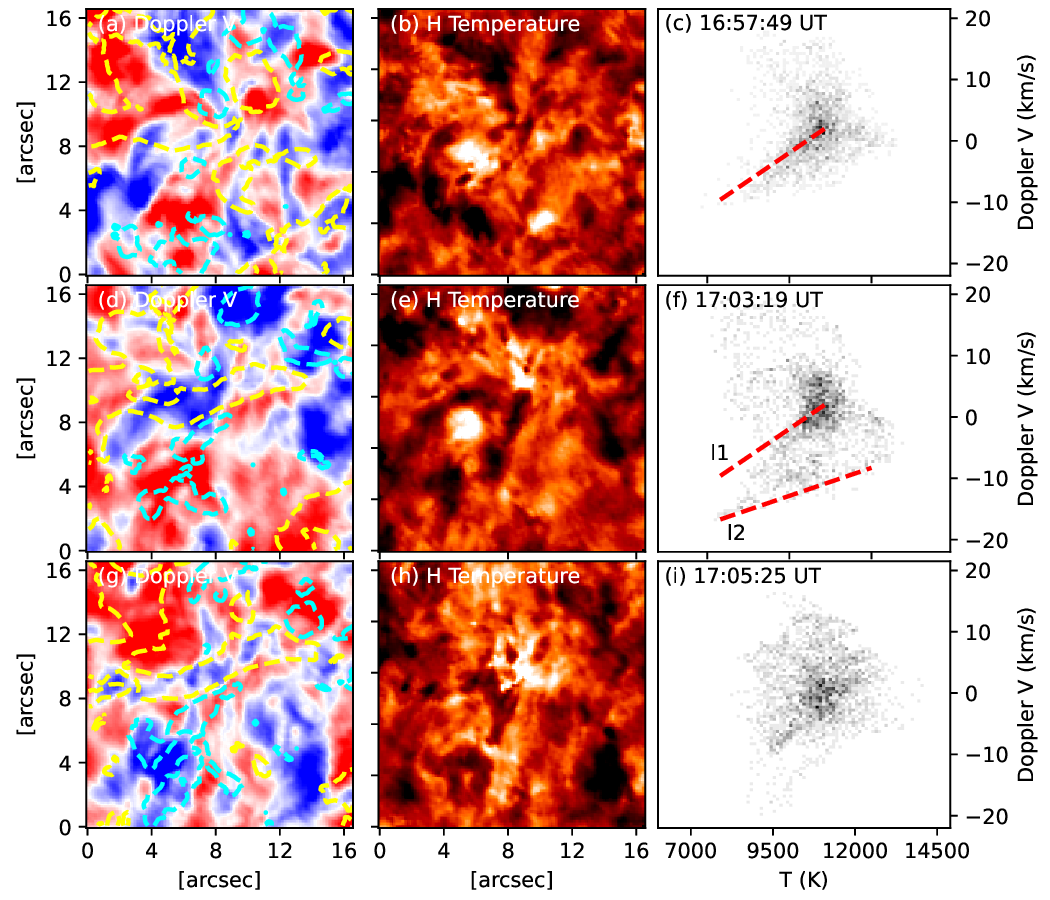}
    \caption{Dopplergram and Temperature correlation of the MF event. (a), (d), and (g) show Dopplergrams at MF activation (T1, 16:57:49~UT), start time of eruption (T2, 17:03:19~UT) and post-eruption time(T3, 17:05:25~UT). Yellow (blue) contours indicate the brightening features in \ha line center ($-$0.6\AA\ off central wavelength)  (b), (e), and (h) show Temperature maps at T1, T2, and T3, respectively. (c), (f), and (i) show correspondence of Temperature and Doppler velocity of the MF event at T1, T2, and T3, respectively.}
    \label{fig:scatter}
\end{figure}

At [8'',8''] of the Dopplergrams((a),(d), and (g)) and Temperature maps ((b), (e), and (h)) in figure \ref{fig:scatter}, prominent upflow is shown along the MF sigmoid associated with temperature burst near the central region. 
The scatterplots at activation time 16:57:49~UT in Fig.\ref{fig:scatter}(c) show temperature in the MF region is \sm9,700~K with cool materials of 7,500~K rising at Doppler speed of 10~\kms. Then with enhanced blue shift observed at start of eruption in the Dopplergram (Fig.\ref{fig:scatter}(d)), a segment of high speed area (l2) that over 10~\kms\ appear in the scattered distribution, while the temperature does not show promiment change.
At post-eruption time 17:05:25~UT, the scatterplot (Fig.\ref{fig:scatter}(i)) shows 2D Gaussian like distribution centered at 9,800~K. We observed that with upflow speed over 10~\kms, temperature is 9,500~K, which is 2,000~K higher than start of the MF eruption.
\subsection{EUV brightenings associated with the MF eruption}
\begin{figure}[ht!]
    \centering
    \includegraphics[width=\textwidth]{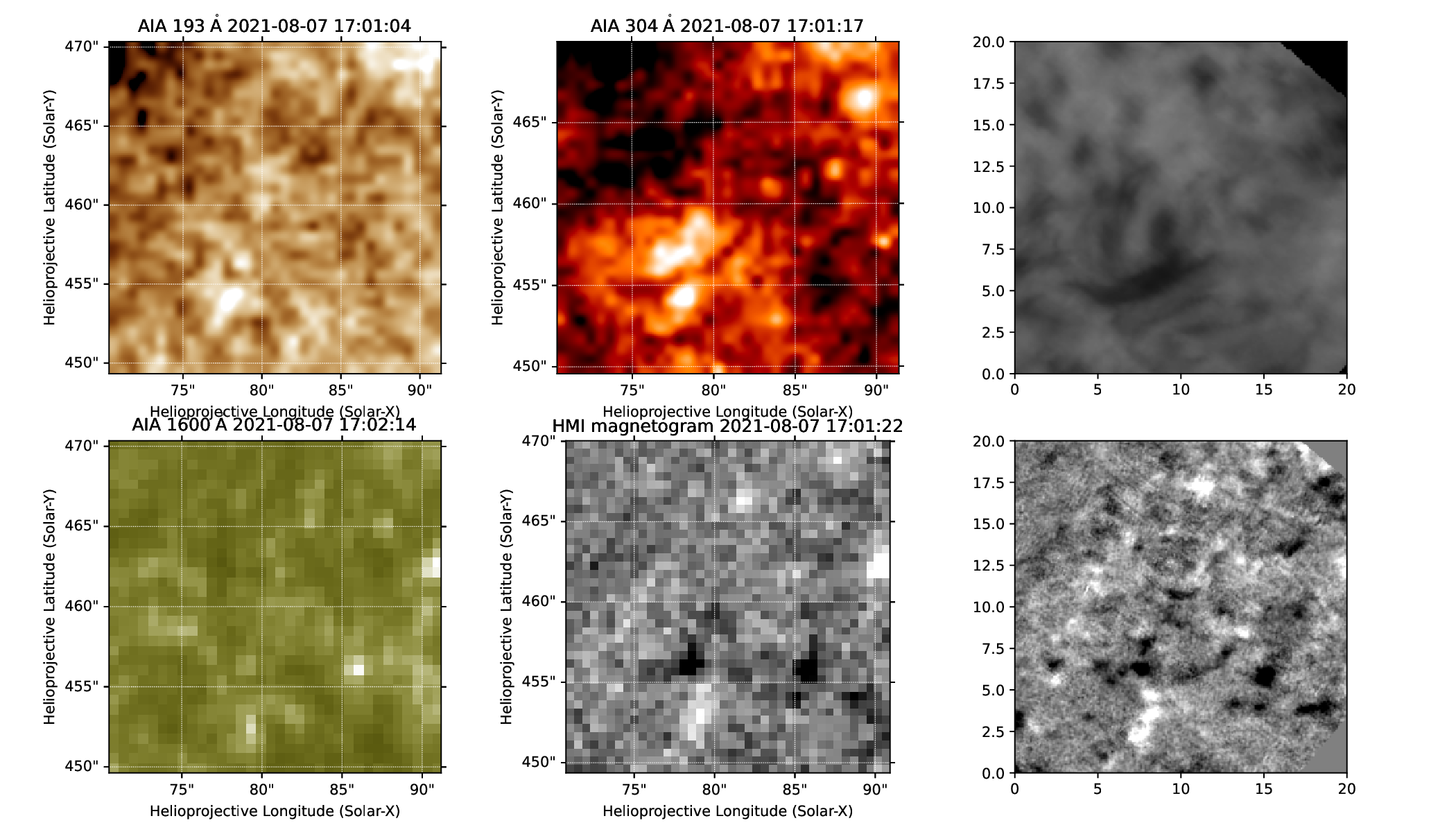}
    \caption{MF eruption in UV/EUV images and magnetograms at 17:01~UT. Panel (a), (b) and (d) show AIA images in 193, 304, and 1600~\AA\ channels. }
    \label{fig:euv_brights}
\end{figure}

\begin{figure}
    \centering
    \includegraphics[width=\textwidth]{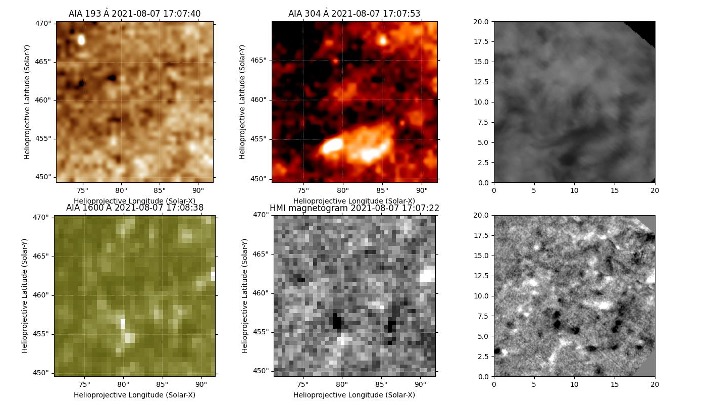}
    \caption{MF eruption in UV/EUV images and magnetograms at 17:07~UT. Panel (a), (b) and (d) show AIA images in 193, 304, and 1600~\AA\ channels. }
    \label{fig:euv_ribbon}
\end{figure}
 
Figure \ref{fig:euv_brights} and Figure \ref{fig:euv_ribbon} show EUV brightenings in AIA 193~\AA\ (Figure \ref{fig:euv_brights} (a)) with corresponding MF at 17:01:04~UT (Figure \ref{fig:euv_brights} (b) and two-ribbon flare at 17:07:53~UT (Figure \ref{fig:euv_ribbon} (b)). The EUV brightenings occur as the dark filament feature appear from bright background in AIA 304~\AA, and vanishes when two-ribbon flaring configuration is observed in chromosphere. \cite{2020ApJ...902....8C} found that at onset of the MF, it has no corresponding brightening signals and experiences slow-to-fast acceleration. while in this event, not only the acceleration is associated with appearance of rising filament, brightening is also observed at coronal heights before flare ribbon appearance. This indicates reconnection of a blowout jet model, in which the interchange reconnection with external field and reconnection of internal flux loop occurred the same time. 

To estimate the energy distribution of the MF event, we applied the cloud model assumption proposed by \cite{1997A&A...324.1183T} to determine plasma density in the chromospheric cloud features. This allows us to calculate kinetic and thermal energy release of the MF eruption from fitted thermodynamic properties obtained in Section \ref{sub1}. Combining AIA 304~\AA\ and \ha\ line-center images, we find the eruptive filament of 12~\arcsec length and maximum width of 2\farcs5. Adopting the mass density estimation of 1.1$\times$$10^{-13}$~$cm~s^{-3}$, the kinetic energy change of the rising filament is 1.50$\times$$10^{25}$~ergs, and thermal energy accumulation in the MF is 1.44$\times$$10^{25}$~ergs. From the photospheric magnetograms, we find the magnetic energy change is 1.57$\times$$10^{26}$~ergs across the PIL of converging opposite magnetic elements. Our previous study of quiet Sun small-scale eruptions suggest that magnetic flux cancellation in granule scale provides sufficient free energy to heat the corona \citep{Wang2022}. 

\section{Summary and Discussions} \label{sec:conclusions}

In this paper, we investigate the chromospheric thermodynamics of minifilament eruption associated with brightening and corresponding photospheric magnetic cancellation with high-cadence spectral data of \ha\ line from FISS, and high-resolution magnetograms from NIRIS of GST. With updated MLSI of \ha\ data, temperature evolution and energy transportation of chromespheric features are quantitatively evaluated at a small-scale brightening region.  The coronal bright points are observed in AIA EUV (193, 304~\AA) channel imaging data as micro-flare in the quiet Sun, which is associated with partial minifilament eruption.  
Key structural and dynamic properties of the eruption minifilament are summarized as follows.
\begin{enumerate}

    \item With high-resolution spectroscopy,   we analyzed the dynamic process of this very small scale sigmoidal mini-filament.  Such micro-flare related minifilament eruption is not associated with jet/jetlet in vicinity, while coronal brightening is observed along the trajectory direction of the eruption. This  resemblances two-ribbon flare with filament eruption in the large scale.

    \item Temperature rise in the flaring regions is co-temporal with magnetic flux emergence while the brightening emission in corona is associated with photospheric flux cancellation. Time evolution of the temperature in the filament sigmoid show maximum value when lateral displacement is observed in the flux loop. Temperature estimation shows increase from 8,800~K to 12,600~K, an increase of almost  4,000~K during the eruptive phase, while the maximum Doppler velocity (18~\kms) of the filament is observed at beginning of the eruption.
    
    \item Thermal and kinetic energy releases are both in the order of $1.5\times10^{25}$ ergs,  consist of 18.7\% of total magnetic energy change in the photosphere from emergence phase to flux cancellation. The released energy is considered as minimum of minifilament eruption release, which is comparable to enhanced spicular events. 
\end{enumerate}


Minifilament eruptions are known as downscale filament activities in the quiet Sun. Systematic survey of 32 days of BBSO \ha\ data for small-scale eruptive filamentary events suggest such ubiquitous presence as many as 600 per 24 hour day \citep{1986NASCP2442..369H}. Aside from topological similarities between minifilament events and large scale filaments that often has one or both ends anchor at cancellation site, small-scale filament also undergo emergence phase (slow rising phase) and eruption phase, which are often associated with microflares.
We report observations of microflare ribbons in \ha\ that are often unresolved due to resolution limit of instrument. The microflare is visible in AIA 304~\AA\ as well, with j-shaped dark feature overlying along the PIL of the magnetic bipole in the photosphere. In the eruptive phase, the sigmoid feature experienced heating processing and lateral displacement. Such chromospheric heating can be interpreted as results of microflare triggered particle beams \citep{Frogner_2020}.

Typically, both cool and hot plasma are both ejected from eruptions. In this event, we have not observed cool plasma ejection into the corona that are often associated with (mini)filament eruption and (micro)flare. \cite{Sterling2022} proposed a small-scale analog of coronal jet, in which the cool materials of erupting minifilamet fill in the open field jet spire. However, not all coronal jets are associated with presence of erupting cool minifilament. \cite{Kumar_2019} found 33\% of the coronal jets in their study have no evidence of cool plasma at base of the spire. One argument of the missing coronal jet is that such small-scale eruptive activity have weak signal in EUV, especially cool spire can be smeared by background signals. The observed filament is of typical small-scale filament size, with coronal brightening counterparts of the magnetic cancellation site. 
Study of coronal jet eruptions by \cite{Panesar2020} conclude that magnetic cancellation triggers the eruption before runaway breakout/tether-cutting reconnection when speed of the rising minifilament is above 1~\kms. Our results show that the minifilament experience explosive motion in vertical direction with speed over 10~\kms\ following magnetic flux cancellation in photosphere, which suggest the runaway reconnection at the internal current sheet drove minifilament eruption. However, transition from slow rise to fast rise (eruption) is not observed. Doppler velocity decreases when lateral expansion of the minifilament is observed in \ha\ images. Coincident of magnetic flux cancellation and acceleration of minifilament eruption in this study agree with previous studies of large filament eruptions that tether-cutting reconnection accelerate eruptions \citep{2006GMS...165...43M}.

With high cadence \ha\ spectral data observed by FISS, we found that the minifilament in the chromosphere has close connection with CBP phenomena. In the event of mini-filament eruption, jet is not associated with the chromospheric eruptive motions as identified in previous observations. Instead, the loop shape brightening in corona become visible following the minifilamentary eruption. 
Analysis of magnetic properties from NIRIS data indicates that the cancellation of a magnetic bipole in the footpoints of minifilament triggered the mini-eruption. Line-of-sight velocity and temperature from an inversion of \ha\ spectra with embedded cloud model exhibit a temperature increase of 13,300 K in the brightening  region, associated with average minifilament rising speed increased by 21~$km~s^{-1}$ in 10~minutes. Such a study of high-resolution observations provides further details of connections between photospheric magnetic topology and its chromospheric counterpart, with which we conclude that energy released through mini-eruptions could be a source of coronal heating.
\begin{acknowledgments} 
We gratefully acknowledge the use of data from the Goode Solar Telescope (GST) of the Big Bear Solar Observatory (BBSO). BBSO operation is supported by US NSF AGS-2309939 and AGS-1821294 grants and New Jersey Institute of Technology. GST operation is partly supported by the Korea Astronomy and Space Science Institute and the Seoul National University. This work was supported by NSF grants, AGS-2114201, AGS-2229064 and AGS-2309939. It is also supported by NASA grants, 80NSSC19K0257, 80NSSC20K1282 and 80NSSC24K0258. 
\end{acknowledgments}

\vspace{5mm}
\facilities{GST(FISS and NIRIS), SDO(AIA)}



\bibliography{MyRefs}
\bibliographystyle{aasjournal}



\end{document}